\documentclass[12pt,twoside]{article}

\usepackage[a4paper]{geometry}
\makeatletter
\renewcommand\title[1]{\gdef\@title{\reset@font\Large\bfseries #1}}
\renewcommand\section{\@startsection {section}{1}{\z@}%
                                   {-3.5ex \@plus -1ex \@minus -.2ex}%
                                   {2.3ex \@plus.2ex}%
                                   {\normalfont\large\bfseries}}
\renewcommand\subsection{\@startsection{subsection}{2}{\z@}%
                                     {-3ex\@plus -1ex \@minus -.2ex}%
                                     {1.5ex \@plus .2ex}%
                                     {\normalfont\normalsize\bfseries}}
\renewcommand\subsubsection{\@startsection{subsubsection}{3}{\z@}%
                                     {-2.5ex\@plus -1ex \@minus -.2ex}%
                                     {1.5ex \@plus .2ex}%
                                     {\normalfont\normalsize\bfseries}}

\def\@runningauthor{}\newcommand{\runningauthor}[1]{\def\runningauthor{#1}}
\def\@runningtitle{}\newcommand{\runningtitle}[1]{\def\runningtitle{#1}}

\g@addto@macro\bfseries{\boldmath}

\makeatother

\usepackage{amsthm,amsmath,amssymb}
\usepackage{graphicx}
\usepackage[colorlinks=true,citecolor=black,linkcolor=black,urlcolor=blue]{hyperref}
\theoremstyle{plain}
\newtheorem{theorem}{Theorem}
\newtheorem{lemma}{Lemma}

\theoremstyle{definition}
\newtheorem{definition}{Definition}

\theoremstyle{remark}

\usepackage{subeqnarray}
\usepackage{cases}
\usepackage{amsmath}
\allowdisplaybreaks[4]
\usepackage{amssymb}

\title{Several new infinite classes of 0-APN power functions over $\mathbb{F}_{2^n}$}

\author{\small Yuying Man$^a$, Shizhu Tian$^b$\footnote{Corresponding author. Email addresses: Shizhutian@hubu.edu.cn}, Nian Li$^b$, Xiangyong Zeng$^a$
\\
\footnotesize $^a$ Hubei Key Laboratory of Applied Mathematics, Faculty of Mathematics and Statistics, \\
\footnotesize Hubei University, Wuhan 430062, China\\
\footnotesize $^b$ Hubei Key Laboratory of Applied Mathematics, School of Cyber Science and Technology, \\
\footnotesize Hubei University, Wuhan 430062, China
}

\date{}
\newcommand{\gf}{{\mathbb F}}

\begin{document}
\maketitle
\thispagestyle{empty}

\begin{abstract}
The investigation of partially APN functions has attracted a lot of research interest recently. In this paper, we present several new infinite classes of 0-APN power functions over $\mathbb{F}_{2^n}$ by using the
multivariate method and resultant elimination, and show that these 0-APN power functions are CCZ-inequivalent to the known ones.
\vspace{2mm}

\end{abstract}


\section{Introduction}

\hspace*{0.4cm} Let $\mathbb{F}_{2^n}$ be the finite field with $2^n$ elements and $\mathbb{F}_{2^n}^*=\gf_{2^n}\setminus \{0\}$, where $n$ is a positive integer.
Let $F(x)$ be a mapping from  $\mathbb{F}_{2^n}$ to itself. The derivative function, denoted by $\mathbb{D}_aF$, of $F(x)$ at an element $a$ in $\gf_{2^n}$ is given  by
$\mathbb{D}_aF(x)=F(x+a)-F(x).$
For any $a,\,b \in \gf_{2^n}$,  let  $\delta_F(a,b)=|\{x \in \gf_{2^n}~| ~\mathbb{D}_{a}F(x)=b\}|,$
where $|S|$ denotes the cardinality of a set $S$,
and define
$\delta(F)=\max \{ \delta_F(a,b)~| ~a \in \mathbb{F}_{2^n}^*,\,\, b \in \mathbb{F}_{2^n}\}.$
Nyberg defined a mapping $F(x)$ to be \emph{differentially $\delta$-uniform} iff $\delta(F)=\delta$ \cite{Nyberg1994}, and $\delta(F)$ is called the \emph{differential uniformity} of $F(x)$ accordingly. The differential uniformity is an important concept in cryptography since it quantifies the security of the mappings which are used  in many block ciphers.
For practical applications in cryptography, it is usually desirable to employ mappings with differential uniformity no greater than 4. For example, the AES (Advanced Encryption Standard) uses the inverse function $x \mapsto x^{-1}$ over $\gf_{2^n}$, which has differential uniformity $4$ for even $n$ and $2$ for odd $n$.

Functions $F(x)$ with differential uniformity $\delta(F)=2$ are called almost perfect nonlinear (APN) functions. APN functions are of great interest due to their importance in the design of S-boxes in block ciphers and their close connection to optimal objects in coding theory and combinatorial theory.
Constructing APN functions has been intensively studied in the last three decades, and by far we only found six classes of APN power functions over $\mathbb{F}_{2^n}$: Gold functions \cite{Gold}, Welch functions \cite{Welch}, Inverse functions \cite{Nyberg1994}, Kasami functions \cite{Kasami}, Dobbertin functions \cite{Dobbertin}, Niho functions \cite{Niho}.

In order to study the conjecture of the highest possible algebraic degree of APN functions, Budaghyan et al. in \cite{BKK} proposed the concept of the partially APN as follows.

\begin{definition}\label{def1}(\cite{BKK1})
Let $F(x)$ be a function from $\mathbb{F}_{2^n}$ to itself. For a fixed $x_0 \in \gf_{2^n}$, the function $F(x)$ is called $x_0$-APN (or partially APN) if all the points $x$, $y$ satisfying $F(x_0)+F(x)+F(y)+F(x_0+x+y)=0$ belong to the curve $(x_0+x)(x_0+y)(x+y)=0$.
\end{definition}

If $F(x)$ is an APN function, it is clear that $F(x)$ is $x_0$-APN for any $x_0 \in \gf_{2^n}$. Budaghyan et al. in \cite{BKK} provided some propositions and characterizations of partially APN functions. Pott in \cite{Pott} noted that there exist partially 0-APN permutations on $\mathbb{F}_{2^n}$ for any $n>3$. When $F(x)$ is a power function, i.e., $F(x)=x^d$ for a positive integer $d$, due to their special algebraic structure, we only need to consider the partial APN properties of $F(x)$ at $x_0=0$ or $x_0=1$. Moreover, $F(x)$ is 0-APN
if and only if the equation $F(x+1)+F(x)+1=0$ has no solution in $\mathbb{F}_{2^n} \setminus \{0,1\}$. Budaghyan et al. listed all power functions over $\gf_{2^n}$ for $1\le n\le 11$ that are 0-APN but not APN power functions in \cite[Table 1]{BKK1}, and they also constructed some classes of 0-APN but not APN power functions over $\gf_{2^n}$ in \cite{BKK, BKK1}. Recently, Qu and Li constructed seven infinite classes of 0-APN power functions over $\gf_{2^n}$ and  one of them was proved to be locally-APN \cite{QL}. Very recently, some infinite classes of 0-APN power functions over $\mathbb{F}_{2^n}$ were constructed in \cite{FY} and \cite{WZ}.

In this paper, we propose several new infinite classes of 0-APN power functions over $\mathbb{F}_{2^n}$ by using the multivariate method and resultant elimination. It is worthy noting that these new infinite classes of 0-APN power functions can cover some examples for $1\le n \le 11$ in \cite[Table 1]{BKK1} which are not explained before. We list our constructed new infinite classes of 0-APN power functions in Table 1. Magma experiments show that all these 0-APN power functions over $\mathbb{F}_{2^n}$ in this paper are CCZ-inequivalent to the known ones.

The remainder of this paper is organized as follows.  Section \ref{pre}  gives some basic results that will be needed in this paper. In Section \ref{main-result1}, we present several infinite classes of 0-APN power functions over finite fields, and the concluding remarks are given in Section \ref{con-remarks}.

\begin{table}[t]\label{table-1}
\caption{New classes of 0-APN power functions $F(x)=x^d$ over $\gf_{2^n}$}
\begin{tabular}{|c|c|c|c|c|}
		\hline
       No. & $d$ & Condition & Ref. &  Values of ($d$, $n$) \\  \hline
       1 & $2^{m+1}+3$ & $n=2m+1$ &  \rm Thm. \ref{thmn=2m+1-1} &  (35, 9) (67, 11) \\  \hline
       2 & $5\cdot 2^m+3$ & $n=2m+1$ & \rm Thm. \ref{thmn=2m+1-1} & (83, 9) (163, 11)\\  \hline
       3 & $3(2^m-1)$ & $n=3m-1$ & $\rm Thm. \ \ref{thmn=3m-1-1}$ & (45, 11) \\  \hline
       4 & $5\cdot 2^{m-1}+1$ & $n=3m-1$, $m\not \equiv 5\,\, ({\rm mod}\,\,14)$ & $\rm Thm. \ \ref{thmn=3m-1-1}$ & (41, 11) \\  \hline
       5 & $2^{2m+1}-3\cdot 2^{m-1}+1$ & $n=3m$, $m\not \equiv 2\,\, ({\rm mod}\,\,3)$  & \rm Thm. \ref{thmn=3m} & (117, 9)\\  \hline
       6 & $2^{2m}+2^{m-1}+1$ & $n=3m+1$  & \rm Thm. \ref{thmn=3m+1-1} & (69, 10)\\  \hline
       7 & $2^{2m}+3\cdot 2^{m-1}-1$ & $n=3m+1$ & \rm Thm. \ref{thmn=3m+1-1} & (75, 10) \\  \hline
       8 & $2^{2m-1}+2^m+1$ & $n=4m-1$ & \rm Thm.  \ref{thmn=4m-1-1} & (41, 11)\\ \hline
       9 & $3\cdot 2^m+1$ & $n=4m-1$ & \rm Thm. \ref{thmn=4m-1-1} & (25, 11)\\ \hline
       10 & $2^{2m-1}-2^{m-1}-1$ & $n=4m-1$ & $\rm Thm.  \ \ref{thmn=4m-1-1}$ & (27, 11)\\ \hline
       11 & $3(2^{2m+1}-1)$ & $n=4m-1$ & \rm Thm. \ref{thmn=4m-1-1}  & (381, 11)\\ \hline
       12 & $2^{2m+1}+2^{m-1}+1$ & $n=4m+1$, $m\not \equiv 13\,\, ({\rm mod}\,\,53)$ & \rm Thm.  \ref{thmn=4m+1-1} & (35, 9)\\ \hline
       13 & $2^{3m}+2^m+1$ & $n=5m$ & $\rm Thm.  \ \ref{thmn=5m-1}$ &(69, 10)\\  \hline
       14 & $2^{2m+1}-2^m-1$ & $n=5m$, $m\not \equiv 0\,\, ({\rm mod}\,\,3)$ & $\rm Thm. \ \ref{thmn=5m-1}$ & (27, 10)\\  \hline	
\end{tabular}	
\footnotesize In the last column, ($d$, $n$) denotes the examples of 0-APN but not APN power functions $x^d$ over $\gf_{2^n}$ ($1\le n \le 11$), which appeared in \cite[Table 1]{BKK1} but not yet explained.
\end{table}


\section{Preliminaries}\label{pre}
 \quad The following lemma will be used frequently in this paper.

\begin{lemma}\label{lemma1-root}\rm (\cite[Thm. 2.14]{FF})
	Let $q$ be a prime power and let $f(x)$ be an irreducible polynomial over $\mathbb{F}_q$ of degree $n$. Then $f(x)=0$ has $n$ distinct roots $x$ in $\mathbb{F}_{q^n}$.
\end{lemma}
In order to prove our main results in this paper, we need give some basic facts about the resultant conclusions of two polynomials.
\begin{definition}\label{res}(\cite{FF})
Let $q$ be a prime power, and $\mathbb{F}_q[x]$ be the polynomial ring over $\mathbb{F}_q$. Let $f(x)=a_0x^n+a_1x^{n-1}+\cdots +a_n\in \mathbb{F}_q[x]$ and $g(x)=b_0x^m+b_1x^{m-1}+\cdots +b_m\in \mathbb{F}_q[x]$ be two polynomials of degree $n$ and $m$ respectively, where $n, m\in \mathbb{N}$. Then the resultant $Res(f,g)$ of $f(x)$ and $g(x)$ is defined by the determinant
$$ {\small \begin{array}{c@{\hspace{-5pt}}l}
Res(f, g)=\begin{vmatrix}
    a_{0} & a_{1} & \cdots & a_{n} & 0 & & \cdots & 0 \\
  0 & a_{0} & a_{1} & \cdots & a_{n} & 0 & \cdots & 0 \\
  \vdots & & & & & & & \vdots \\
  0 & \cdots & 0 & a_{0} & a_{1} & & \cdots & a_{n} \\
  b_{0} & b_{1} & \cdots & & b_{m} & 0 & \cdots & 0 \\
  0 & b_{0} & b_{1} & \cdots & & b_{m} & \cdots & 0 \\
 \vdots & & & & & & & \vdots \\
  0 & \cdots & 0 & b_{0} & b_{1} & & \cdots & b_{m}
  \end{vmatrix}
\begin{array}{l}
 \left.\rule{0mm}{11mm}\right\}m\,\rm rows \\
 \\
 \left.\rule{0mm}{11mm}\right\}n\,\rm rows
 \end{array}
 \\ [-5pt]
\end{array}}$$
of order $m+n$.
\end{definition}

If $deg(f)=n$ (i.e., if $a_0 \ne 0$) and $f(x)=a_0(x-\alpha_1)\cdots (x-\alpha_n)$ in the splitting field of $f$ over $\mathbb{F}_q$, then $Res(f,g)$ is also given by the formula
$$Res(f,g)=a_0^m\prod_{i=1}^{n} g(\alpha_i).$$
In this case, we obviously have $Res(f, g)=0$ if and only if $f$ and $g$ have a common root, which is the same as saying that $f$ and $g$ have a common divisor in $\mathbb{F}_q[x]$ of a positive degree.

For two polynomials $F(x, y), G(x, y)\in \mathbb{F}_q[x, y]$ of positive degrees in $y$, the resultant $Res(F,G,y)$ of $F$ and $G$ with respect to $y$ is the resultant of $F$ and $G$ when considered as polynomials in the univariate $y$. In this case, $Res(F, G, y)\in \mathbb{F}_q[x]\cap \langle F,G\rangle$, where $\langle F,G\rangle$ is the ideal generated by $F$ and $G$. Thus any pair $(a,b)$ with $F(a,b)=G(a,b)=0$ is such that $Res(F,G,y)(a)=0$.


\section{New classes of 0-APN power functions over $\mathbb{F}_{2^n}$}\label{main-result1}
\quad In this section, we present several new classes of 0-APN power functions over $\mathbb{F}_{2^n}$ by using the multivariate method and resultant elimination.
\subsection{The case of $n=2m+1$}\label{n=2m+1}
\quad In this subsection, two new classes of 0-APN power functions over $\mathbb{F}_{2^n}$ are given for $n=2m+1$.
\begin{theorem}\label{thmn=2m+1-1}
Let $n$, $m$ be positive integers with $n=2m+1$. Then $F(x)=x^d$ is 0-APN over $\mathbb{F}_{2^n}$ when one of the following statements holds:
\begin{itemize}
\item [\rm 1)] $d=2^{m+1}+3$;
\item [\rm 2)] $d=5\cdot 2^m+3$.
\end{itemize}
\begin{proof}
We only consider the case 1) since the other case can be similarly proved. To complete the proof, it is sufficient to prove that the equation
\begin{equation}\label{2m+1-2-1}
(x+1)^{2^{m+1}+3}+x^{2^{m+1}+3}+1=0
\end{equation}
has no solution in $\mathbb{F}_{2^n}\setminus \{0,1\}$.
Observe that (\ref{2m+1-2-1}) can be reduced to
\begin{equation}\label{2m+1-1-2}
x^{2^{m+1}+2}+x^{2^{m+1}+1}+x^{2^{m+1}}+x^3+x^2+x=0.
\end{equation}
Let $y=x^{2^m}$. Raising the $2^m$-th power to (\ref{2m+1-1-2}), we have
\begin{subequations}
\begin{numcases}{}
y^2x^2+y^2x+y^2+x^3+x^2+x=0, \label{thm2m+1-2-eq-1-1}\\
xy^2+xy+x+y^3+y^2+y=0. \label{thm2m+1-2-eq-1-2}
\end{numcases}
\end{subequations}
We compute the resultant of (\ref{thm2m+1-2-eq-1-1}) and (\ref{thm2m+1-2-eq-1-2}) with respect to $y$ (which exactly means the resultant of the left hand sides of (\ref{thm2m+1-2-eq-1-1}) and (\ref{thm2m+1-2-eq-1-2})), and the resultant can be decomposed into the following product of irreducible factors in $\gf_{2}$ as
$$x(x+1)(x^2+x+1)^4.$$
Note that $x\ne 0, 1$. Suppose that $x^2+x+1=0$. According to Lemma \ref{lemma1-root}, we have $x\in \gf_{2^2}$, however, $ \gf_{2^2}\cap \gf_{2^n}=\gf_{2}$, which contradicts with $x\not \in \gf_{2}$. Therefore, (\ref{2m+1-2-1}) has no solution in $\mathbb{F}_{2^n}\setminus \{0,1\}$. This completes the proof.
\end{proof}
\end{theorem}


\subsection{The case of $n=3m-1$}\label{n=3m-1}
\quad In this subsection, we present two  new classes of 0-APN power functions over $\mathbb{F}_{2^n}$ of $n=3m-1$.
\begin{theorem}\label{thmn=3m-1-1}
Let $n$, $m$ be positive integers with $n=3m-1$. Then $F(x)=x^d$ is 0-APN over $\mathbb{F}_{2^n}$ when one of the following statements holds:
\begin{itemize}
\item [\rm 1)] $d=5\cdot 2^{m-1}+1$, where $m\not \equiv 5\,\, ({\rm mod}\,\,14)$;
\item [\rm 2)] $d=3(2^m-1)$.
\end{itemize}
\begin{proof}
We only prove the case 1) as the other case can be similarly proved. We next show that the equation
\begin{equation}\label{3m-1-2-1}
(x+1)^{5\cdot 2^{m-1}+1}+x^{5\cdot 2^{m-1}+1}+1=0
\end{equation}
has no solution in $\mathbb{F}_{2^n}\setminus \{0,1\}$. By a direct calculation, (\ref{3m-1-2-1}) can be written as
\begin{equation}\label{3m-1-2-2}
x^{5\cdot 2^{m-1}}+x^{4\cdot 2^{m-1}+1}+x^{4\cdot 2^{m-1}}+x^{2^{m-1}+1}+x^{2^{m-1}}+x=0.
\end{equation}
Squaring both sides of (\ref{3m-1-2-2}) gives
\begin{equation}\label{3m-1-2-3}
x^{5\cdot 2^m}+x^{4\cdot 2^m+2}+x^{4\cdot 2^m}+x^{2^m+2}+x^{2^m}+x^2=0.
\end{equation}
Let $y=x^{2^m}$ and $z=y^{2^m}$. Raising the $2^m$-th power and $2^{2m}$-th power to equation (\ref{3m-1-2-3}) respectively obtains
\begin{subequations}
\begin{numcases}{}
y^5+y^4x^2+y^4+yx^2+y+x^2=0, \label{thm3m-1-2-eq-1-1}\\
z^5+z^4y^2+z^4+zy^2+z+y^2=0, \label{thm3m-1-2-eq-1-2}\\
x^{10}+x^8z^2+x^8+x^2z^2+x^2+z^2=0. \label{thm3m-1-2-eq-1-3}
\end{numcases}
\end{subequations}
Computing the resultant of (\ref{thm3m-1-2-eq-1-2}) and (\ref{thm3m-1-2-eq-1-3}) with respect to $z$, we have
$$Res(x, y)=(x^{25} + x^{24}y^2 + x^{17}y^2 + x^{17} + x^{16} + x^9y^2 + x^8y^2 + x^8 + x + y^2)^2.$$
Next, we compute the resultant of (\ref{thm3m-1-2-eq-1-1}) and $Res(x, y)$ with respect to $y$, by Magma computation, the resultant can be decomposed into the product of irreducible factors in $\gf_{2}$. In addition to $x$ and $x+1$, the degree of each of these irreducible factors is 7, which implies that $x\in \gf_{2^7}$ by Lemma \ref{lemma1-root}. Since $n=3m-1$ and $m\not \equiv 5\,\, ({\rm mod}\,\,14)$, it can be verified that $\gcd(7, n)=1$. This leads to $x\in \gf_{2^7}\cap \gf_{2^n}=\gf_{2}$, a contradiction. It follows that (\ref{3m-1-2-1}) has no solution in $\mathbb{F}_{2^n}\setminus \{0,1\}$. This completes the proof.
\end{proof}
\end{theorem}


\subsection{The case of $n=3m$}\label{n=3m}
\quad In this subsection, we present a  new class of 0-APN power function over $\mathbb{F}_{2^n}$ of $n=3m$.
\begin{theorem}\label{thmn=3m}Let $n$, $m$ be positive integers  with $m\not \equiv 2\,\, ({\rm mod}\,\,3)$ and $n=3m$. Then
$F(x)=x^{2^{2m+1}-3\cdot 2^{m-1}+1}$
is a 0-APN function over $\mathbb{F}_{2^n}$.
\begin{proof}
We show that the equation
\begin{equation}\label{3m-1}
(x+1)^{2^{2m+1}-3\cdot 2^{m-1}+1}+x^{2^{2m+1}-3\cdot 2^{m-1}+1}+1=0
\end{equation}
has no solution in $\mathbb{F}_{2^n}\setminus \{0,1\}$. Multiplying $x^{3\cdot 2^{m-1}}(x+1)^{3\cdot 2^{m-1}}$ on both sides of (\ref{3m-1}), then (\ref{3m-1}) can be written as
\begin{equation}\label{3m-2}
\begin{split}
&x^{2^{2m+1}+3\cdot 2^{m-1}}+x^{3\cdot 2^{m-1}+1}+x^{2^{2m+1}+2^m+1}+x^{2^{2m+1}+2^{m-1}+1}\\
&+x^{2^{2m+1}+1}+x^{3\cdot 2^m}+x^{5\cdot 2^{m-1}}+x^{2^{m+1}}=0.
\end{split}
\end{equation}
Raising the square to (\ref{3m-2}) leads to
\begin{equation}\label{3m-3}
\begin{split}
&x^{2^{2m+2}+3\cdot 2^m}+x^{3\cdot 2^m+2}+x^{2^{2m+2}+2^{m+1}+2}+x^{2^{2m+2}+2^m+2}\\
&+x^{2^{2m+2}+2}+x^{3\cdot 2^{m+1}}+x^{5\cdot 2^m}+x^{2^{m+2}}=0.
\end{split}
\end{equation}
Let $y=x^{2^m}$ and $z=y^{2^m}$. Raising the $2^m$-th power and $2^{2m}$-th power to equation (\ref{3m-3}) respectively gives
\begin{subequations}
\begin{numcases}{}
z^4y^3+y^3x^2+z^4y^2x^2+z^4yx^2+z^4x^2+y^6+y^5+y^4=0, \label{thm3m-eq-1}\\
x^4z^3+z^3y^2+x^4z^2y^2+x^4zy^2+x^4y^2+z^6+z^5+z^4=0, \label{thm3m-eq-2}\\
y^4x^3+x^3z^2+y^4x^2z^2+y^4xz^2+y^4z^2+x^6+x^5+x^4=0. \label{thm3m-eq-3}
\end{numcases}
\end{subequations}
Computing the resultants of (\ref{thm3m-eq-1}) and (\ref{thm3m-eq-2}), (\ref{thm3m-eq-1}) and (\ref{thm3m-eq-3}) with respect to $z$ respectively, we have
\begin{equation*}
\begin{split}
Res_1(x, y)=&y^8(y+1)^8(x^2+y)^4(x^{20}y^6 + x^{20}y^5 + x^{20}y^3 + x^{20}y + x^{20} + x^{18}y^8 + x^{18}y^2 \\
&+ x^{16}y^{12}
+x^{16}y^{10} + x^{16}y^9 + x^{16}y^8 + x^{16}y^7 + x^{16}y^4 + x^{16}y^3 + x^{14}y^8 + x^{14}y^4 \\
&+ x^{12}y^9 + x^{12}y^8
+ x^{12}y^5 + x^{12}y^4 +
 x^{10}y^{10} + x^{10}y^6 + x^8y^{12} + x^8y^{11} + x^8y^8 \\
 &+ x^8y^7 + x^6y^{12} + x^6y^8
 + x^4y^{13} + x^4y^{12} + x^4y^9 + x^4y^8 +
 x^4y^7 + x^4y^6 + x^4y^4 \\
 &+ x^2y^{14} + x^2y^8 + y^{16} + y^{15}
 + y^{13} + y^{11} + y^{10}),
\end{split}
\end{equation*}
\begin{equation*}
\begin{split}
Res_2(x, y)=&(x+y)^8(x^{10}y^2 + x^{10}y + x^{10} + x^8y^3 + x^8y^2 + x^8y + x^8 + x^6y^6 + x^6y^5
+ x^6y^4 \\&+ x^6y^3 +
x^6y^2 + x^6y + x^6 + x^4y^{10} + x^4y^9 + x^4y^8 + x^4y^7 + x^4y^6 + x^4y^5 + x^4y^4 \\
&+ x^2y^{10} + x^2y^9 + x^2y^8 + x^2y^7 +
y^{10} + y^9 + y^8)^2.
\end{split}
\end{equation*}
Next, we compute the resultant of $Res_1(x, y)$ and $Res_2(x, y)$ with respect to $y$, with the help of Magma, the resultant can be decomposed into the following product of irreducible factors in $\gf_{2}$ as
\begin{equation}\label{3m-res}
\begin{split}
&x^{480}(x+1)^{480}(x^2+x+1)^{64}(x^3+x^2+1)^{48}(x^3+x+1)^{48}
(x^{12} + x^{11}
+ x^8 + x^6 \\&+ x^4 + x^3 + x^2 + x + 1)^4(x^{12} + x^{11} + x^{10} + x^9 + x^8 + x^6 + x^4 + x + 1)^4.
\end{split}
\end{equation}
Observe that $x\not \in \gf_{2}$, thus the solutions of (\ref{3m-res}) are in $\gf_{2^2}$, $\gf_{2^3}$ or $\gf_{2^{12}}$ by Lemma \ref{lemma1-root}. Next we show that $x\not \in \gf_{2^2}$, $x\not \in \gf_{2^3}$ and $x\not \in \gf_{2^{12}}$.

\rm (1)
If $x\in \gf_{2^2}$. When $m$ is odd, we have $x\in \gf_{2^2}\cap \gf_{2^n}=\gf_{2}$, which is a contradiction. When $m$ is even, we have $x\in \gf_{2^2}\cap \gf_{2^n}=\gf_{2^2}$. Then $x^{2^m}=x$ and $x^{2^{2m}}=x$. Thus we derive from (\ref{3m-3}) that $x^2+x=0$, which contradicts with $x\ne 0, 1$.

\rm (2)
If $x\in \gf_{2^3}$. When $m\equiv 1\,\, ({\rm mod}\,\,3)$, we get $x^{2^m}=x^2$ and $x^{2^{2m}}=x^4$. It follows from (\ref{3m-3}) that
$$x^6+x^5+x^4+x^3=x^3(x+1)^3=0,$$
it is impossible since $x\ne 0, 1$. When $m\equiv 0\,\, ({\rm mod}\,\,3)$, we obtain $x^{2^m}=x$ and $x^{2^{2m}}=x$. Hence, we derive from (\ref{3m-3}) that $x^4+x=x(x+1)(x^2+x+1)=0$. Recall that $x\not \in \gf_{2^2}$, a contradiction.

\rm (3)
If $x\in \gf_{2^{12}}$. When $m\equiv 1\,\, ({\rm mod}\,\,4)$ or $m\equiv 3\,\, ({\rm mod}\,\,4)$, we have $x\in \gf_{2^{12}}\cap \gf_{2^n}=\gf_{2^3}$, which contradicts with the above discussion.

When $m\equiv 2\,\, ({\rm mod}\,\,4)$, we can get $x\in \gf_{2^{12}}\cap \gf_{2^n}=\gf_{2^6}$. If $m\equiv 1\,\, ({\rm mod}\,\,3)$, then we obtain $x^{2^m}=x^{16}$ and $x^{2^{2m}}=x^4$. It follows from (\ref{3m-3}) that
 $$x^{34}+x^{33}+x^{18}+x^{17}=x^{17}(x+1)^{17}=0,$$
which means $x\in \gf_{2}$. It leads to a contradiction. If $m\equiv 0\,\, ({\rm mod}\,\,3)$, we get $x^{2^m}=x$ and $x^{2^{2m}}=x$, we conclude from (\ref{3m-3}) that $x^8+x^4=x^4(x+1)^4=0$, thus $x\in \gf_{2}$. It contradicts with $x\ne 0, 1$.

When $m\equiv 0\,\, ({\rm mod}\,\,4)$, we have $x\in \gf_{2^{12}}\cap \gf_{2^n}=\gf_{2^{12}}$. If $m\equiv 1\,\, ({\rm mod}\,\,3)$, we obtain $x^{2^m}=x^{16}$ and $x^{2^{2m}}=x^{256}$. Hence, we derive from (\ref{3m-3}) that
$$x^{1072}+x^{1058}+x^{1042}+x^{1026}+x^{96}+x^{80}+x^{64}+x^{50}=0.$$
We can decompose the above equation into the product of irreducible factors in $\gf_{2}$. Apart from $x$ and $x+1$, the degree of each of these irreducible factors is in the set of $\{8, 54, 150\}$. Then we have $x\in \gf_{2^8}$, $x\in \gf_{2^{54}}$ or $x\in \gf_{2^{150}}$ by Lemma \ref{lemma1-root}. Observe that $\gf_{2^8}\cap \gf_{2^{12}}=\gf_{2^4}$. Then $x\in \gf_{2^4}$, we have $x^{2^m}=x$ and $x^{2^{2m}}=x$, thus we derive from (\ref{3m-3}) that $x^8+x^4=x^4(x+1)^4=0$, which contradicts with $x\ne 0, 1$. When $x\in \gf_{2^{54}}$ and $x\in \gf_{2^{150}}$, we have $\gf_{2^{54}}\cap \gf_{2^{12}}=\gf_{2^{150}}\cap \gf_{2^{12}}=\gf_{2^6}$, then $x\in \gf_{2^6}$. We have $x^{2^m}=x^{16}$ and $x^{2^{2m}}=x^4$. It follows from (\ref{3m-3}) that
$x^{34}+x^{33}+x^{18}+x^{17}=x^{17}(x+1)^{17}=0,$
which is a contradiction. If $m\equiv 0\,\, ({\rm mod}\,\,3)$, we get $x^{2^m}=x$ and $x^{2^{2m}}=x$, we conclude from (\ref{3m-3}) that $x^8+x^4=x^4(x+1)^4=0$, which is impossible due to $x\ne 0, 1$.
 Hence, (\ref{3m-1}) has no solution in $\mathbb{F}_{2^n}\setminus \{0,1\}$. The proof is completed.
\end{proof}
\end{theorem}


\subsection{The case of $n=3m+1$}\label{n=3m+1}
\quad In this subsection, two new classes of 0-APN power functions over $\mathbb{F}_{2^n}$ are given for $n=3m+1$.
\begin{theorem}\label{thmn=3m+1-1}
Let $n$, $m$ be positive integers with $n=3m+1$. Then $F(x)=x^d$ is 0-APN over $\mathbb{F}_{2^n}$ when one of the following statements holds:
\begin{itemize}
\item [\rm 1)] $d=2^{2m}+2^{m-1}+1$;
\item [\rm 2)] $d=2^{2m}+3\cdot 2^{m-1}-1$.
\end{itemize}
\begin{proof}
It suffices to show that the equation
\begin{equation}\label{3m+1-1-0}
F(x+1)+F(x)+1=0
\end{equation}
has no solution in $\mathbb{F}_{2^n}\setminus \{0,1\}$.

1) When $d=2^{2m}+2^{m-1}+1$, note that (\ref{3m+1-1-0}) is equivalent to
\begin{equation}\label{3m+1-1-2}
x^{2^{2m}+2^{m-1}}+x^{2^{2m}+1}+x^{2^{2m}}+x^{2^{m-1}+1}+x^{2^{m-1}}+x=0.
\end{equation}
Squaring both sides of (\ref{3m+1-1-2}) gives
\begin{equation}\label{3m+1-1-3}
x^{2^{2m+1}+2^m}+x^{2^{2m+1}+2}+x^{2^{2m+1}}+x^{2^m+2}+x^{2^m}+x^2=0.
\end{equation}
Let $y=x^{2^m}$ and $z=y^{2^m}$. Taking the $2^m$-th power on equation (\ref{3m+1-1-3}) derives
\begin{subequations}
\begin{numcases}{}
z^2y+z^2x^2+z^2+yx^2+y+x^2=0, \label{thm3m+1-eq-1-1}\\
xz+xy^2+x+zy^2+z+y^2=0. \label{thm3m+1-eq-1-2}
\end{numcases}
\end{subequations}
Computing the resultant of (\ref{thm3m+1-eq-1-1}) and (\ref{thm3m+1-eq-1-2}) with respect to $z$, and the resultant can be decomposed into the product of irreducible factors in $\gf_{2}$ as
$$y(y+1)(x^2+x+1)^2(y^2+y+1).$$
If the above equation equals zero, we have $x^2+x+1=0$ or $y^2+y+1=0$ since $y\not \in \gf_{2}$.

Suppose that $x^2+x+1=0$. If $m$ is even, then $x\in \gf_{2^2}\cap \gf_{2^n}=\gf_{2}$, a contradiction. If $m$ is odd, let $\omega\in \gf_{2^2}\setminus \{0,1\}$, then $\omega$, $\omega^2$  are  solutions of $x^2+x+1=0$. Plugging $x=\omega$ into (\ref{3m+1-1-2}) gives $1=0$, a contradiction. Similarly, we can also prove $y^2+y+1\ne 0$. Thereby (\ref{3m+1-1-2}) has no solution in $\mathbb{F}_{2^n}\setminus \{0,1\}$. This completes the proof of 1).

2) When $d=2^{2m}+3\cdot 2^{m-1}-1$, multiplying $x(x+1)$ on both sides of (\ref{3m+1-1-0}), then
(\ref{3m+1-1-0}) can be written as
\begin{equation}\label{3m+1-2-2}
x^{2^{2m}+3\cdot 2^{m-1}}+x^{3\cdot 2^{m-1}+1}+x^{2^{2m}+1}+x^{2^{2m}+2^m+1}+x^{2^{2m}+2^{m-1}+1}+x^{2^m+1}+x^{2^{m-1}+1}+x^2=0.
\end{equation}
Raising the square to (\ref{3m+1-2-2}) results in
\begin{equation}\label{3m+1-2-3}
\begin{split}
&x^{2^{2m+1}+3\cdot 2^m}+x^{3\cdot 2^m+2}+x^{2^{2m+1}+2}+x^{2^{2m+1}+2^{m+1}+2} \\
&+x^{2^{2m+1}+2^m+2}+x^{2^{m+1}+2}+x^{2^m+2}+x^4=0.
\end{split}
\end{equation}
Let $y=x^{2^m}$ and $z=y^{2^m}$. Raising the $2^m$-th power and $2^{2m+1}$-th power to equation (\ref{3m+1-2-3}) respectively, we obtain
\begin{subequations}
\begin{numcases}{}
z^2y^3+y^3x^2+z^2x^2+z^2y^2x^2+z^2yx^2+y^2x^2+yx^2+x^4=0, \label{thm3m+1-2-eq-1-1}\\
xz^3+z^3y^2+xy^2+xz^2y^2+xzy^2+z^2y^2+zy^2+y^4=0, \label{thm3m+1-2-eq-1-2}\\
y^2x^3+x^3z^4+y^2z^4+y^2x^2z^4+y^2xz^4+x^2z^4+xz^4+z^8=0. \label{thm3m+1-2-eq-1-3}
\end{numcases}
\end{subequations}
Computing the resultants of (\ref{thm3m+1-2-eq-1-1}) and (\ref{thm3m+1-2-eq-1-2}), (\ref{thm3m+1-2-eq-1-1}) and (\ref{thm3m+1-2-eq-1-3}) with respect to $z$ respectively gets
\begin{equation*}
\begin{split}
Res_1(x,y)=&(x+y)^2(x^{12} + x^{10}y^6 + x^{10}y^5 + x^{10}y^3 + x^{10}y + x^8y^7 + x^8y^5 + x^8y^4 + x^8y^3\\
&+x^8y^2 + x^6y^{12} + x^6y^{10} + x^6y^5 + x^6y^3 + x^4y^{13} + x^4y^{12} + x^4y^{11} + x^4y^{10} + x^4y^8 \\
&
+x^2y^{14} +
x^2y^{12} + x^2y^{10} + x^2y^9 + y^{15}),
\end{split}
\end{equation*}
\begin{equation*}
\begin{split}
Res_2(x,y)=&x^6(x+1)^6(x+y^2)^2(x^9 + x^8y^4 + x^7y^4 + x^7y^2 + x^6y^6 + x^6y^4 + x^5y^8 + x^5y^6 \\
 &+x^5y^4 + x^5y^2 + x^4y^{10} + x^4y^8 + x^4y^6 + x^4y^4 + x^3y^8 + x^3y^6 + x^2y^{10} + x^2y^8 \\
 &+ xy^8 +y^{12})^2.
\end{split}
\end{equation*}
Computing the resultant of $Res_1(x,y)$ and $Res_2(x,y)$ with respect to $y$, and the resultant can be decomposed into the product of irreducible factors in $\gf_{2}$ as
$$x^{346}(x+1)^{346}.$$
Observe that $x\not \in \gf_{2}$. Thus (\ref{3m+1-2-2}) has no solution in $\mathbb{F}_{2^n}\setminus \{0,1\}$. This completes the proof of 2).
\end{proof}
\end{theorem}


\subsection{The case of $n=4m-1$}\label{n=4m-1}
\quad In this subsection, we present four  new classes of 0-APN power functions over $\mathbb{F}_{2^n}$ of $n=4m-1$.
\begin{theorem}\label{thmn=4m-1-1}
Let $n$, $m$ be positive integers with $n=4m-1$. Then $F(x)=x^d$ is 0-APN over $\mathbb{F}_{2^n}$ when one of the following statements holds:
\begin{itemize}
\item [\rm 1)] $d=3\cdot 2^m+1$;
\item [\rm 2)] $d=2^{2m-1}+2^m+1$;
\item [\rm 3)] $d=2^{2m-1}-2^{m-1}-1$;
\item [\rm 4)] $d=3(2^{2m+1}-1)$.
\end{itemize}
\begin{proof}
We only consider the cases 1) and 2) since others can be similarly proved. In order to complete the proof, we will prove that the equation
\begin{equation}\label{4m-1-0}
F(x+1)+F(x)+1=0
\end{equation}
has no solution in $\mathbb{F}_{2^n}\setminus \{0,1\}$.

1) When $d=3\cdot 2^m+1$. Observe that (\ref{4m-1-0}) can be reduced to
\begin{equation}\label{4m-1-3-2}
x^{3\cdot 2^{m}}+x^{2^{m+1}+1}+x^{2^{m+1}}+x^{2^m+1}+x^{2^m}+x=0.
\end{equation}
Let $y=x^{2^m}$, $z=y^{2^m}$ and $u=z^{2^m}$. Taking $2^m$-th power, $2^{2m}$-th power and $2^{3m}$-th power on the equation (\ref{4m-1-3-2}) respectively, we obtain
\begin{subequations}
\begin{numcases}{}
y^3+y^2x+y^2+yx+y+x=0, \label{thm4m-1-3-eq-1-1}\\
z^3+z^2y+z^2+zy+z+y=0, \label{thm4m-1-3-eq-1-2}\\
u^3+u^2z+u^2+uz+u+z=0, \label{thm4m-1-3-eq-1-3}\\
x^6+x^4u+x^4+x^2u+x^2+u=0. \label{thm4m-1-3-eq-1-4}
\end{numcases}
\end{subequations}
Computing the resultant of (\ref{thm4m-1-3-eq-1-3}) and (\ref{thm4m-1-3-eq-1-4}) with respect to $u$ gives
$$Res_1(x, z)=(x^2+z)(x^2+x+1)^8.$$
Next, we compute the resultant of (\ref{thm4m-1-3-eq-1-2}) and $Res_1(x, z)$ with respect to $z$ as follows
$$Res_2(x, y)=(x^2+y)(x^2+x+1)^{26}.$$
Finally, computing the resultant of $(\ref{thm4m-1-3-eq-1-1})$ and $Res_2(x, y)$ with respect to $y$, and decomposing the resultant into the following product of irreducible factors in $\gf_{2}$ as
$$x(x+1)(x^2+x+1)^{80}.$$
If the above equation equals zero, we have $x^2+x+1=0$ due to $x\ne 0, 1$, thus $x\in \gf_{2^2}$. However, $\gf_{2^2}\cap \gf_{2^n}=\gf_{2}$, which is a contradiction. It follows that (\ref{4m-1-3-2}) has no solution in $\mathbb{F}_{2^n}\setminus \{0,1\}$. This completes the proof of 1).

2) When $d=2^{2m-1}+2^m+1$. Notice that (\ref{4m-1-0}) can be simplified as
\begin{equation}\label{4m-1-1-2}
x^{2^{2m-1}+2^m}+x^{2^{2m-1}+1}+x^{2^{2m-1}}+x^{2^m+1}+x^{2^m}+x=0.
\end{equation}
Squaring both sides of (\ref{4m-1-1-2}) gives
\begin{equation}\label{4m-1-1-3}
x^{2^{2m}+2^{m+1}}+x^{2^{2m}+2}+x^{2^{2m}}+x^{2^{m+1}+2}+x^{2^{m+1}}+x^2=0.
\end{equation}
Let $y=x^{2^m}$, $z=y^{2^m}$ and $u=z^{2^m}$. Taking the $2^m$-th power, $2^{2m}$-th power and $2^{3m}$-th power on the equation (\ref{4m-1-1-3}) respectively, we obtain
\begin{subequations}
\begin{numcases}{}
zy^2+zx^2+z+y^2x^2+y^2+x^2=0, \label{thm4m-1-eq-1-1}\\
uz^2+uy^2+u+z^2y^2+z^2+y^2=0, \label{thm4m-1-eq-1-2}\\
x^2u^2+x^2z^2+x^2+u^2z^2+u^2+z^2=0, \label{thm4m-1-eq-1-3}\\
y^2x^4+y^2u^2+y^2+x^4u^2+x^4+u^2=0. \label{thm4m-1-eq-1-4}
\end{numcases}
\end{subequations}
Computing the resultants of (\ref{thm4m-1-eq-1-2}) and (\ref{thm4m-1-eq-1-3}), (\ref{thm4m-1-eq-1-2}) and (\ref{thm4m-1-eq-1-4}) with respect to $u$ respectively gives
$$Res_1(x,y,z)=(y^2xz^2+y^2xz+y^2z^3+y^2z^2+y^2+xz^3+xz+x+z^2+z)^2,$$
$$Res_2(x,y,z)=(y^3x^2+y^3z^2+y^2x^2z^2+y^2z^2+y^2+yx^2z^2+yx^2+y+x^2+z^2)^2.$$
Next computing the resultants of (\ref{thm4m-1-eq-1-1}) and $Res_1(x,y,z)$, (\ref{thm4m-1-eq-1-1}) and $Res_2(x,y,z)$ with respect to $z$ respectively obtains
$$Res_3(x,y)=(y^8x^6+y^8x^5+y^8x^3+y^8x^2+y^8+x^7+x^5+x^4+x^2+x)^2,$$
$$Res_4(x,y)=(x^2+x+1)^4(y^7+y^6x^2+y^5x^2+y^5+y^4+y^3x^2+y^2x^2+y^2+y+x^2)^2.$$
Computing the resultant of $Res_3(x,y)$ and $Res_4(x,y)$ with respect to $y$, the resultant can be decomposed into the product of some irreducible factors in $\gf_{2}$ as
\begin{equation}\label{4m+1-res}
\begin{split}
&x^4(x+1)^4(x^2+x+1)^{128}(x^5+x^2+1)^4(x^5+x^3+1)^4(x^5+x^3+x^2+x+1)^4\\
&(x^5+x^4+x^2+x+1)^4(x^5+x^4+x^3+x+1)^4(x^5+x^4+x^3+x^2+1)^4.
\end{split}
\end{equation}
 Notice that $x\ne 0, 1$, now we suppose that the resultant of $Res_3(x,y)$ and $Res_4(x,y)$ with respect to $y$ equals zero, then we obtain $x\in \gf_{2^2}$ or $x\in \gf_{2^5}$ by Lemma \ref{lemma1-root}. When $x\in \gf_{2^2}$, we have $x\in \gf_{2^2}\cap \gf_{2^n}=\gf_{2}$ since $n=4m-1$, a contradiction. When $x\in \gf_{2^5}$, then $x^{2^m}=x^{16}$ and $x^{2^{2m}}=x^8$ if $m\equiv 4\,\, ({\rm mod}\,\,5)$. We derive from (\ref{4m-1-1-3}) that $$x^{10}+x^9+x^8+x^3+x^2+x=x(x+1)(x^2+x+1)(x^3+x+1)(x^3+x^2+1)=0.$$
 Recall that $x\not \in \gf_{2^2}$. We have $x^2+x+1\ne 0$. Suppose one of the  equations $x^3+x+1=0$ and $x^3+x^2+1=0$ holds, then $x\in \gf_{2^3}$. However, $x\in \gf_{2^3}\cap \gf_{2^5}=\gf_{2}$, which contradicts with $x\ne 0, 1$. If $n=4m-1$ and $m\not \equiv 4\,\, ({\rm mod}\,\,5)$, we have $\gcd(5,n)=1$ and $x\in \gf_{2^5}\cap \gf_{2^n}=\gf_{2}$. This is impossible since $x\ne 0, 1$. Therefore (\ref{4m-1-1-2}) has no solution in $\mathbb{F}_{2^n}\setminus \{0,1\}$. This completes the proof of 2).
\end{proof}
\end{theorem}


\subsection{The case of $n=4m+1$}\label{n=4m+1}
\quad In this subsection, a new class of 0-APN power function over $\mathbb{F}_{2^n}$ is given for $n=4m+1$.
\begin{theorem}\label{thmn=4m+1-1}
Let $n$ and $m$ be positive integers with $m\not \equiv 13\,\, ({\rm mod}\,\,53)$ and $n=4m+1$. Then
$F(x)=x^{2^{2m+1}+2^{m-1}+1}$
is a 0-APN function over $\mathbb{F}_{2^n}$.
\begin{proof}
It suffices to show that the equation
\begin{equation}\label{4m+1-1}
(x+1)^{2^{2m+1}+2^{m-1}+1}+x^{2^{2m+1}+2^{m-1}+1}+1=0
\end{equation}
has no solution in $\mathbb{F}_{2^n}\setminus \{0,1\}$. Observe that (\ref{4m+1-1}) can be simplified as
\begin{equation}\label{4m+1-2}
x^{2^{2m+1}+2^{m-1}}+x^{2^{2m+1}+1}+x^{2^{2m+1}}+x^{2^{m-1}+1}+x^{2^{m-1}}+x=0.
\end{equation}
Raising the square to (\ref{4m+1-2}) leads to
\begin{equation}\label{4m+1-3}
x^{2^{2m+2}+2^m}+x^{2^{2m+2}+2}+x^{2^{2m+2}}+x^{2^m+2}+x^{2^m}+x^2=0.
\end{equation}
Let $y=x^{2^m}$, $z=y^{2^m}$ and $u=z^{2^m}$. Raising the $2^m$-th power, $2^{2m+1}$-th power and $2^{3m+1}$-th power to the equation (\ref{4m+1-3}) respectively, we obtain
\begin{subequations}
\begin{numcases}{}
z^4y+z^4x^2+z^4+yx^2+y+x^2=0, \label{thm4m+1-eq-2-1}\\
u^4z+u^4y^2+u^4+zy^2+z+y^2=0, \label{thm4m+1-eq-2-2}\\
x^4u^2+x^4z^4+x^4+u^2z^4+u^2+z^4=0, \label{thm4m+1-eq-2-3}\\
y^4x+y^4u^4+y^4+xu^4+x+u^4=0. \label{thm4m+1-eq-2-4}
\end{numcases}
\end{subequations}
Computing the resultants of (\ref{thm4m+1-eq-2-2}) and (\ref{thm4m+1-eq-2-3}), (\ref{thm4m+1-eq-2-2}) and (\ref{thm4m+1-eq-2-4}) with respect to $u$ respectively gives
$$Res_1(x,y,z)=(x^8y^2z^8+x^8y^2z+x^8z^9+x^8z^8+x^8+y^2z^9+y^2z+y^2+z^8+z)^2,$$
$$Res_2(x,y,z)=(xy^6+xy^4z+xy^4+xy^2z+x+y^6z+y^4+y^2z+y^2+z)^4.$$
Next computing the resultants of (\ref{thm4m+1-eq-2-1}) and $Res_1(x,y,z)$, (\ref{thm4m+1-eq-2-1}) and $Res_2(x,y,z)$ with respect to $z$ respectively obtains
\begin{equation*}
\begin{split}
Res_1(x,y)=&(x^{50}y^{16}+x^{50}y+x^{50}+x^{48}y^{17}+x^{48}y^{16}+x^{48}y+x^{34}y^{17}+x^{34}+x^{32}y^{16}+x^{32}y\\
&+x^{18}y^{17}+x^{18}y^{16}+x^{18}y+x^{16}y^{17}+x^{16}+x^2y^{16}+x^2y+x^2+y^{17}+y^{16}+y)^2,
\end{split}
\end{equation*}
$$Res_2(x,y)=(x^2+x+1)(x^2y^{24}+x^2y^{17}+x^2y^9+x^2y^8+x^2+y^{25}+y^{17}+y^{16}+y^8+y)^4.$$
Assume that $x^2+x+1=0$. We have $x\in \mathbb{F}_{2^2}$ by Lemma \ref{lemma1-root}. Thus $x\in
\mathbb{F}_{2^2}\cap \mathbb{F}_{2^n}=\mathbb{F}_{2}$ for $n=4m+1$, which is a contradiction. Hence $x^2+x+1\ne 0$. Computing the resultant of $Res_1(x,y)$ and $Res_2(x,y)/(x^2+x+1)$ with respect to $y$, and then by Magma computation, the resultant can be decomposed into the product of irreducible factors in $\gf_{2}$. In addition to $x$ and $x+1$, the degree of each of these irreducible factors is 53, it leads to $x\in \gf_{2^{53}}$ by Lemma \ref{lemma1-root}. When $n=4m+1$, we have $\gcd(53, n)=1$ or $\gcd(53, n)=53$, and $(4m+1)\equiv 0\,\, ({\rm mod}\,\,53)$ if and only if $m\equiv 13\,\, ({\rm mod}\,\,53)$. Thus if $m\not \equiv 13\,\, ({\rm mod}\,\,53)$, we have $\gcd(53, n)=1$. Then we have $x\in \gf_{2^{53}}\cap \gf_{2^n}=\gf_{2}$, a contradiction. It follows that (\ref{4m+1-1}) has no solution in $\mathbb{F}_{2^n}\setminus \{0,1\}$. This completes the proof.
\end{proof}
\end{theorem}


\subsection{The case of $n=5m$}\label{n=5m}
\quad In this subsection, we give two new classes of 0-APN power functions over $\mathbb{F}_{2^n}$ for $n=5m$.
\begin{theorem}\label{thmn=5m-1}
Let $n$ and $m$ be  positive integers with $n=5m$. Then $F(x)=x^d$ is 0-APN over $\mathbb{F}_{2^n}$ when one of the following statements holds:
\begin{itemize}
\item [\rm 1)] $d=2^{3m}+2^m+1$;
\item [\rm 2)] $d=2^{2m+1}-2^m-1$, where $m\not \equiv 0\,\, ({\rm mod}\,\,3)$.
\end{itemize}
\begin{proof}
We only consider the case 1) since the other case can be similarly proved. In order to complete the proof, it is sufficient to prove that the equation
\begin{equation}\label{5m1-1}
(x+1)^{2^{3m}+2^m+1}+x^{2^{3m}+2^m+1}+1=0
\end{equation}
has no solution in $\mathbb{F}_{2^n}\setminus \{0,1\}$. Notice that (\ref{5m1-1}) can be written as
\begin{equation}\label{5m1-2}
x^{2^{3m}+2^m}+x^{2^{3m}+1}+x^{2^{3m}}+x^{2^m+1}+x^{2^m}+x=0.
\end{equation}
Let $y=x^{2^m}$, $z=y^{2^m}$, $u=z^{2^m}$ and $v=u^{2^m}$, then $v^{2^m}=x$. Raising the $2^m$-th, $2^{2m}$-th, $2^{3m}$-th and $2^{4m}$-th powers to equation (\ref{5m1-2}) respectively, we have
\begin{subequations}
\begin{numcases}{}
uy+ux+u+yx+y+x=0, \label{thm1-eq-2-1}\\
vz+vy+v+zy+z+y=0, \label{thm1-eq-2-2}\\
xu+xz+x+uz+u+z=0, \label{thm1-eq-2-3}\\
yv+yu+y+vu+v+u=0, \label{thm1-eq-2-4}\\
zx+zv+z+xv+x+v=0. \label{thm1-eq-2-5}
\end{numcases}
\end{subequations}
Computing the resultants of (\ref{thm1-eq-2-2}) and (\ref{thm1-eq-2-4}), (\ref{thm1-eq-2-2}) and (\ref{thm1-eq-2-5}) with respect to $v$ respectively gives
$$Res_1(y,z,u)=(z+u)(y^2+y+1),$$
$$Res_2(x,y,z)=(z^2+z+1)(y+x).$$
Next, computing the resultants of (\ref{thm1-eq-2-1}) and $Res_1(y,z,u)$, (\ref{thm1-eq-2-3}) and $Res_1(y,z,u)$ with respect to $u$ respectively obtains
$$Res_3(x,y,z)=(yx+yz+y+xz+x+z)(y^2+y+1),$$
$$Res_4(x,y,z)=(x+z^2)(y^2+y+1).$$
Assume that $y^2+y+1=0$. We have $(y^2+y+1)^{2^{4m}}=x^2+x+1=0$, then $x\in \mathbb{F}_{2^2}$. When $m$ is odd, we get $x\in
\mathbb{F}_{2^2}\cap \mathbb{F}_{2^n}=\mathbb{F}_{2}$, which is a contradiction. When $m$ is even, we have $x\in
\mathbb{F}_{2^2}\cap \mathbb{F}_{2^n}=\mathbb{F}_{2^2}$, then $x^{2^m}=x$ and $x^{2^{3m}}=x$. Thus we derive from (\ref{5m1-2}) that $x^2+x=0$, which is impossible since $x\ne 0, 1$. Hence $y^2+y+1\ne 0$.
Computing the resultants of $Res_3(x,y,z)/(y^2+y+1)$ and $Res_4(x,y,z)/(y^2+y+1)$, $Res_3(x,y,z)/(y^2+y+1)$ and $Res_2(x,y,z)$ with respect to $z$ respectively, one gets
$$Res_1(x,y)=(x^2+x+1)(y^2+x),$$
$$Res_2(x,y)=(x^2+x+1)(y+x)(y^2+y+1).$$
Recall that $x^2+x+1\ne 0$ and $y^2+y+1\ne 0$, we compute the resultant of $Res_1(x,y)/(x^2+x+1)$ and $Res_2(x,y)/((x^2+x+1)(y^2+y+1))$ with respect to $y$, we have $x(x+1)$. If $x(x+1)=0$, which contradicts with $x\ne 0,1$. Therefore, (\ref{5m1-1}) has no solution in $\gf_{2^n}\setminus \{0,1\}$. The proof is completed.
\end{proof}
\end{theorem}

\section{Conclusion}\label{con-remarks}
\quad In this paper, we mainly presented several new infinite classes of 0-APN power functions over $\gf_{2^n}$ by using the multivarite method and resultant elimination. According to the result on CCZ-equivalence of power functions given in \cite[Thm. 1]{CCZ}, Magma experiments showed that all these 0-APN power functions over $\gf_{2^n}$ in this paper are CCZ-inequivalent to the known ones and our results are CCZ-inequivalent to each other.
\vspace{3mm}



\begin{thebibliography}{999}

   \bibitem{BKK}L. Budaghyan, N. Kaleyski, S. Kwon, C. Riera, P. Stanica. Partially APN Boolean functions and classes of functions that are not APN infinitely often, Cryptography Commun. 12 (2020) 527-545.

    \bibitem{BKK1}L. Budaghyan, N. Kaleyski, C. Riera, P. Stanica. Partially APN functions with APN-like polynomial representations, Des. Codes Cryptogr. 88 (2020) 1159-1177.

    \bibitem{CCZ} U. Dempwolff. CCZ equivalence of power functions, Des. Codes Cryptogr. 86 (2018) 665-692.

    \bibitem{Welch} H. Dobbertin. Almost perfect nonlinear power functions on $\gf_{2^n}$: the Welch case,  IEEE Trans. Inf. Theory 45(4) (1999) 1271-1275.

    \bibitem{Niho} H. Dobbertin. Almost perfect nonlinear power functions on $\gf_{2^n}$: the Niho case, IEEE Trans. Inf. Theory 151 (1999) 57-72.

    \bibitem{Dobbertin} H. Dobbertin. Almost perfect nonlinear power functions on $\gf_{2^n}$: a new case for $n$ divisible by 5, In: Proceedings of Finite Fields and Applications, Augsburg, Germany, (1999) 113-121.

    \bibitem{FY}T. Fu, H. Yan. Several classes of 0-APN power functions over $\mathbb{F}_{2^n}$, arXiv:2210.15103 (2022).

    \bibitem{Gold} R. Gold. Maximal recursive sequences with 3-valued recursive cross-correlation functions, IEEE Trans. Inf. Theory 14 (1968) 154-156.

    \bibitem{Kasami} T. Kasami. The weight enumerators for several classes of subcodes of the 2nd order binary Reed-Muller codes, Information and Control 18(4) (1971) 369-394.

    \bibitem{FF}R. Lidl, H. Niederreiter. Finite fields, Encyclopedia of Mathematics, Cambridge, UK.: Cambridge University Press (1997).

    \bibitem{Nyberg1994}K. Nyberg. Differentially uniform mappings for cryptography, Advances in cryptology - EUROCRYPT'93. Lecture Notes in Computer Science 76 (1994) 55-64.

    \bibitem{Pott}A. Pott. Partially almost perfect nonlinear permutations, In LOOPS, Hungary (2019).

    \bibitem{QL}L. Qu, K. Li. More infinite classes of APN-like power functions, arXiv:2209.13456 (2022).

    \bibitem{WZ}Y. Wang, Z. Zha. New results of 0-APN power functions over $\mathbb{F}_{2^n}$, arXiv:2210.02207 (2022).
\end{thebibliography}
\end{document}